\documentclass[prb,preprint,longbibliography]{revtex4-1} 

\usepackage{bm}
\usepackage{helvet}
\usepackage{amsmath}  
\usepackage{amsfonts} 
\usepackage{graphicx} 

\def\rd{{\rm d}}
\def\re{{\rm e}}
\def\ri{{\rm i}}

\def\tfrac#1#2{\displaystyle \frac{#1}{#2}}
\def\lbar{\lambda\hskip-6.0pt\vrule height5.5pt depth-5.0pt width6pt}

\begin{document}

\title{On the dimension of angles and their units}

\author{Peter J. Mohr}
\email{mohr@nist.gov}
\affiliation{National Institute of Standards and Technology,
Gaithersburg, Maryland 20899}
\author{William D. Phillips}
\email{william.phillips@nist.gov}
\affiliation{Joint Quantum Institute,
University of Maryland, College Park, Maryland 20742}
\affiliation{National Institute of Standards and Technology
Gaithersburg, Maryland 20899}
\author{Eric Shirley}
\email{eric.shirley@nist.gov}
\affiliation{National Institute of Standards and Technology,
Gaithersburg, Maryland 20899}
\author{Michael Trott}
\email{trott@wolfram.com}
\affiliation{Wolfram Inc.,
100 Trade Center Drive
Champaign, IL 61820-7237}

\begin{abstract}

We examine implications of angles having their own dimension, in the
same sense as do lengths, masses, {\it etc.}  The conventional practice
in scientific applications involving trigonometric or exponential
functions of angles is to assume that the argument is the numerical part
of the angle when expressed in units of radians.  It is also assumed
that the functions are the corresponding radian-based versions.  These
(usually unstated) assumptions generally allow one to treat angles as if
they had no dimension and no units, an approach that sometimes leads to
serious difficulties.  Here we consider arbitrary units for angles and
the corresponding generalizations of the trigonometric and exponential
functions. Such generalizations make the functions complete, that is,
independent of any particular choice of unit for angles.  They also
provide a consistent framework for including angle units in computer
algebra programs.

\end{abstract}

\date{\today}

\maketitle 

\section{Introduction}

Angle is a familiar concept that needs no formal definition. When two
lines cross in a plane, two pairs of vertical angles are formed. Unless
the lines are perpendicular, one pair of angles is acute and the other
pair is obtuse.  In trigonometry, the geometric properties of interior
angles of right triangles may be used to derive general relations
between various trigonometric functions. For example, the formula for
the sine of the sum of two angles in terms of sine and cosine functions
of each of the angles, does not depend on a quantitative specification
of the angles.  That is, when we write $\sin{(\alpha+\beta)} =
\sin{\alpha}\,\cos{\beta} + \cos{\alpha}\,\sin{\beta}$ we do not specify
whether $\alpha$ and $\beta$ are measured in degrees or radians or
grads, nor do we specify that the appropriate functions be used,
according to the units of the arguments.  This and other features of
angles lead to the often-repeated (and often misunderstood) assertion
that angles do not have units and therefore do not have a dimension (or
that their dimension is unity).

By contrast, when angles appear in physical measurements, it is
essential to specify the unit of measure for the angle.  For example, it
would be of no use to specify the angle subtended by the distance
between two stars, or the angle between two laser beams, or the rotation
angle of a wheel without giving units like arcseconds or milliradians or
full rotations.  

All of these comments about plane angles apply equally to phase angles.
For example, the amplitude of an electromagnetic plane wave of fixed
frequency and propagation vector (wave vector) is proportional to a sine
function of an argument, a phase angle, whose value changes linearly
with time and distance.  If we want to report the difference in phase
angle between two spacetime points, for example, we must specify the
units--degrees or radians or cycles (full rotations).  

Expressions for the frequency of periodic phenomena is a particularly
important and troublesome example of the difficulties associated with
units for angles.  Frequency is the rate of change of phase angle with
time.  As such, it is just as important that the units of frequency
reflect the units of angles as that they express the units of time.
Frequencies are usually given in units of radians per second or hertz,
meaning cycles (full rotations or revolutions of the phase angle) per
second.  Less often, frequencies might be given in degrees per second or
revolutions per minute.  Unfortunately, sometimes frequencies are simply
expressed in inverse seconds, leading to confusion about what the unit
really is.  

If we agree that angles have units, we must also agree that they have a
dimension, the dimension of angle, which is independent of other
dimensions like length, time, or mass.  Sometimes it is argued that
angles do not have an independent dimension, or that the dimension is
unity, because the dimension of angle is length divided by length.  The
justification given for this is that one may measure angles as arc
length divided by radius vector length.  However, this ratio gives only
the numerical factor of the quantity expression for the angle.  The
complete quantity must include a unit, radian in this case, otherwise,
the numerical factor could be the number of degrees of the angle, or
anything else.  Moreover, that is only one way in which angle might be
measured.\cite{2019063}  In fact, the most common way of measuring
angles is with a protractor, whereby unknown angles are compared to
known angles.  This is exactly analogous to measuring an unknown length
by comparing it to known lengths.  Angles and lengths are manifestly
different kinds of physical quantities, and each has its associated
dimension.  

While we argue that angles are physical quantities distinct from other
physical quantities or combinations of such quantities, and that they
therefore have their own dimension and unit, we recognize, as have many
others, that this represents a choice, and that that choice is made in
part for the sake of convenience.  The number of base units in the
metric system (now the SI, the International System of Units) has
changed over time, because of changes in what was considered to be most
useful and convenient.  The scientific community could have decided to
have a unit system in which temperature is measured in joules, but we
find it to be more convenient to measure temperature in kelvins.  In
some sense we have the same kind of choice here with angles, and we
argue that the advantage achieved by giving angles a separate dimension
and unit outweigh any difficulties.

Our examination of the question of how to deal with units for angles is
of interest not only for the important goal of a consistent, unambiguous
use of units in physics, but also for systematically including units for
angles in algebraic computer software applications. The current practice
of sometimes ignoring the units and resolving any resulting ambiguity by
human judgement is not suitable for the digital age.

There is a long history of discussions of the role of units for angles.
\cite{1822001, 1870001, 1873001, 1910001, 1936002, 1961013, 1962014,
1978038, 1978039, 1978040, 1979035, 1979036, 1980036, 1982033, 1985057,
1986059, 1986056, 1986058, 1986057, 1988061, 1991100, 1992087, 1992088,
1992089, 1993136, 1993137, 1997199, 1998167, 2001385, 2002280, 2005364,
2005363, 2005109, 2010207, 2015004, 2015048, 2015049, 2015154, 2016021,
2016022, 2016059, 2016060, 2016023, 2016121, 2017130, 2017154, 2019004,
2019063, 2019108, 2019004, 2019063, 2019108, 2019112, 2019117, 2019121,
2020031, 2020078, 2021025, 2021029} The underlying ideas in this paper
have, in essence, been considered in a multitude of earlier works.
Here, we have synthesized many of those ideas, while paying particular
attention to making the proposed reforms compatible with modern computer
processing.  

While it would be impractical for every argument in a numerical
computation of trigonometric functions to explicitly carry an angle,
having radians, degrees, \dots, in general (compound) units that
describe physical quantities, such as angular velocity or differential
cross sections, are crucial for automated symbolic computations where
results from one computation have to be dimensionally consistent with
the next computation step. For instance, \emph{Mathematica}\cite{math}
considers angles to have an angle dimension. But in the absence of a
well-defined standard, dimensional inconsistencies resulting from
conversions such as ``1 Hz = 1/s'' are unavoidable and currently require
hard-coded heuristics rather than a fully deterministic algorithmic
treatment. The issue of the consistent presence of an angle dimension
becomes even more amplified at the level of equations that involve
physical quantities (see Secs.~\ref{sec:cyclotron} and \ref{sec:cp}
below).

Although there is a diversity of opinions on the question, the
overwhelming majority of papers on the subject acknowledge that angles
should be regarded as having an independent dimension and associated
units.  The difficulty of consistently implementing this proposal is
also often cited as a reason for maintaining the acknowledged
unsatisfactory status quo.

\section{Unit notation}

Following Maxwell \cite{1873001} and the general practice of
international metrology, we specify a physical quantity $Q$ by a
coefficient and a unit as
\begin{eqnarray}
Q = \{Q\}[Q]\,,
\label{eq:unitdef}
\end{eqnarray}
where $\{Q\}$ is a real or complex number as in Maxwell's definition (or
more generally an operator, matrix, {\it etc.}), and $[Q]$
is the unit.
Or to be more precise, we may write
\begin{eqnarray}
Q = \{Q\}_{[Q]}[Q]\,,
\end{eqnarray}
because the value of $\{Q\}$ depends on the unit $[Q]$.

For example, a length of 3 meters, $L = 3$ m, corresponds to $\{L\} = 3$
and $[L] =$ m.  The choice of units in Eq.~(\ref{eq:unitdef}) is not
unique, so we may have
\begin{eqnarray}
Q = \{Q\}_{[Q]_1}[Q]_1 = \{Q\}_{[Q]_2}[Q]_2 .
\label{eq:equiv}
\end{eqnarray}
For the previous example, the length $L$ could (ill-advisedly) also be
expressed in inches as
\begin{eqnarray}
L = 3 \mbox{ m} = 118.11\dots \mbox{ in}\,,
\end{eqnarray}
where $[L]_1 =$ m, $\{L\}_{\rm m}=3$, $[L]_2=$ in, and $\{L\}_{\rm
in}=118.11\dots$.  The quantity $Q$ is the same physical quantity,
regardless of the unit in which it is expressed.

\section{Units for angles}
\label{sec:tfa}

When we consider angles as measurable physical quantities, as with other
such quantities, an angle $\theta$ may be expressed as
\begin{eqnarray}
\theta = \{\theta\}_{[\theta]}[\theta] \, ,
\label{eq:angdef}
\end{eqnarray}
where $[\theta]$ is one of any number of possible units.  The list
includes degree, minute, second, radian, revolution or cycle, grad,
among other possible choices, although those mentioned are commonly
used.  We focus on degree, radian, and revolution ({\it i.e.,} cycle or
period) as examples, with the understanding that generalization to other
units is always possible.

An example of a particular angle, expressed in different units, is
\begin{eqnarray}
\theta &=& 45~{\rm deg}  \qquad \mbox{where } [\theta] = ~{\rm deg}
\mbox{ and }\{\theta\}_{\rm deg} = 45 \, ,
\\[5 pt]
\theta &=& \frac{\pi}{4} \mbox{ rad} 
\qquad \mbox{where } [\theta] = \mbox{rad}
\mbox{ and } \{\theta\}_{\rm rad} = \frac{\pi}{4} \, ,
\\[5 pt]
\theta &=& \frac{1}{8} \mbox{ rev}
\qquad \mbox{where } [\theta] = \mbox{rev}
\mbox{ and } \{\theta\}_{\rm rev} = \frac{1}{8} \, ,
\end{eqnarray}
where the notation 45 deg $\equiv 45^\circ$ is used, and rev is the
abbreviation for revolution.  In these examples, the unit provides
necessary information about the angle.  If an angle were simply given as
a number, say ``$\theta=45$'', with no unit, it could taken to be 45
rad, which is a perfectly respectable angle, although it may not be the
one the author had in mind.

The relationship between plane right triangles and the trigonometric
functions is clear.  For a given interior angle, the sine of this angle
is the ratio of the length of the side opposite the angle to the length
of the hypotenuse.  This is a real number between $0$ and $1$.  (Here we
limit the discussion to real angles, although the extension to complex
angles is considered below.)  Thus the properties of trigonometric
functions are based on geometrical angles, whereas the functions
themselves are usually defined to have numbers as the argument.  This
apparent inconsistency will be addressed in Sec.~\ref{sec:uitf}.

If we are working in an environment where only numbers are being used,
such as a scientific calculator or traditional FORTRAN, then the unit
ambiguity is avoided if it is recognized that the sine function itself
depends on what units are being used.  As is well known, looking up the
sine of a number representing an angle depends on its unit, and vice
versa.   There are different trigonometric tables for angles in degrees
and for angles in radians.  Similarly, when using a calculator to find
the sine or inverse sine of a number, it is necessary to specify which
unit is being assumed for the input or expected as output by, for
example, touching the Deg/Rad key first.  This dependence on the type of
sine function being used can be denoted by writing
\begin{eqnarray}
\sin_{\rm deg}{\left(45\right)} &=& \frac{1}{\sqrt{2}} \, ,
\label{eq:sinedeg}
\\[5 pt]
\sin_{\rm rad}{\left(\frac{\pi}{4}\right)}
&=& \frac{1}{\sqrt{2}} \, ,
\\[5 pt]
\sin_{\rm rev}{\left(\frac{1}{8}\right)} 
&=& \frac{1}{\sqrt{2}} 
\, ,
\end{eqnarray}
where the subscript on the name of the function indicates the unit being
assumed for the angle.  Similarly, for the inverse functions, we have
\begin{eqnarray}
\arcsin_{{\rm deg}}{\left(\frac{1}{\sqrt{2}}\right)} &=& 45 \, ,
\label{eq:sine}
\\[5 pt]
\arcsin_{\rm rad}{\left(\frac{1}{\sqrt{2}}\right)} &=& \frac{\pi}{4}\, ,
\\[5 pt]
\arcsin_{\rm rev}{\left(\frac{1}{\sqrt{2}}\right)} &=& \frac{1}{8}\, ,
\label{eq:invsine}
\end{eqnarray}
where the particular arcsine function being used determines the value
for the inverse.  However, even if the type of arcsine function is
specified, the result on the right-hand side of
Eqs.~(\ref{eq:sine})-(\ref{eq:invsine}) has lost that information.  As a
result, the numbers 45, $\pi/4$, or $1/8$ could be degrees, radians,
revolutions, or anything else.  As humans, we can look back at our notes
and figure it out, but it is useful to retain that information along
with the number, either to provide complete information about the result
or so that it can unambiguously be used for the next step in a series of
calculations.  This is the rationale for using units in the first place,
and it is important for information to be processed by computers.

To accomplish this, specification of the unit is added to the
``output'' of Eqs.~(\ref{eq:sine})-(\ref{eq:invsine}).  Thus for the
angle $\theta = 45$ deg $= \pi/4$ rad $= 1/8$ rev, we have
\begin{eqnarray}
\theta&=& \arcsin_{\rm deg}\left(\frac{1}{\sqrt{2}}\right) \mbox{ deg} 
= 45 \mbox{ deg} \, ,
\label{eq:asdeg}
\\[5 pt]
\theta&=& \arcsin_{\rm rad}\left(\frac{1}{\sqrt{2}}\right) \mbox{ rad} 
= \frac{\pi}{4} \mbox{ rad} \, ,
\\[5 pt]
\theta&=& \arcsin_{\rm rev}\left(\frac{1}{\sqrt{2}}\right) \mbox{ rev} 
= \frac{1}{8} \mbox{ rev} \, .
\label{eq:asrev}
\end{eqnarray}
Note that Eqs.~(\ref{eq:sinedeg})-(\ref{eq:invsine}) are mapping of
numbers to numbers without dimensions, but
Eqs.(\ref{eq:asdeg})-(\ref{eq:asrev}) provide angles with units
included.  These equations may be generalized to an angle
$\theta=\{\theta\}_{[\theta]}[\theta]$ expressed in any unit $[\theta]$,
as in Eq.~(\ref{eq:angdef}), by writing
\begin{eqnarray}
w &=& \sin_{[\theta]}\left(\{\theta\}_{[\theta]}\right)
\label{eq:gensine}
\end{eqnarray}
and
\begin{eqnarray}
\{\theta\}_{[\theta]} &=& \arcsin_{[\theta]}\left(w\right)
\label{eq:geninvsine}
\end{eqnarray}
In general, because the sine function is periodic, an infinite number of
angles map into a particular value of the sine function, so we assume
that the formula for the inverse gives angles in the range:
$|\{\theta\}_{[\theta]}| \le 90,\,\pi/2,\,1/4$, as appropriate.

In this section, we have considered the purely numerical sine and
arcsine functions.  That is, functions that have real numbers for both
their domains and ranges.  The extension to a sine function of angles
with dimensions and any units and to the corresponding arcsine function
that has dimensional angles as its range is examined in the next
section.  Of course, this generalization applies to the cosine and
exponential functions and their inverses as well. 

\section{Unit-independent transcendental functions}
\label{sec:uitf}

Here, we consider the concept of ``complete'' functions.\cite{1914002} 
That is, they are independent of the units in which the arguments are
expressed.  This is a property of trigonometric functions based on plane
geometry, which may be derived with no reference to units for the
angles.  On the other hand, for physical applications, it is useful to
specify units for both the arguments of the functions and for the
functions themselves.  This has been spelled out in detail above for the
three examples of units of  degrees, radians, and revolutions or cycles.
For this purpose, it is useful to make a clear distinction between
trigonometric functions that have angles with dimensions as arguments
and trigonometric functions that have real numbers as arguments, as
considered in the preceding sections.  In the latter case, the real
number is the numerical factor of the angle when expressed in a
particular unit.  The terminology ``geometrical angle'' and ``analytical
angle'' for these differing representations for angles was applied by
\citet{1962014}, although only radian units for the analytic form were
considered at the time.  This terminology has been repeated in many
works since then.  Here we examine the relation between complete
trigonometric functions and the explicit unit forms considered in the
foregoing.

\subsection{Trigonometric functions}

To describe complete transcendental functions, it is necessary to
consider generalizations of their derivatives.  Angles may be
represented as quantities with the dimension of angle, but with an
arbitrary unit.  Infinitesimal changes in an angle correspond to
infinitesimal changes in the ratios of the associated sides of the
triangle that includes the angle.  This relationship provides a
quantitative relation between changes of the angle and changes of the
trigonometric functions.  However, the conventional formula for the
derivative,
\begin{eqnarray}
``\,\frac{\rd}{\rd\theta} \,\sin(\theta) &=& \cos(\theta)\,\mbox{''} \, ,
\end{eqnarray}
is problematic, because the equation is not dimensionally consistent.
Both sine and cosine are dimensionless, as they are ratios of lengths of
the sides of a triangle, so the left-hand side of the equation has the
dimension of the inverse of an angle, and the right-hand side is
dimensionless.  As mentioned previously, angles are not intrinsically
dimensionless.

\begin{figure}
\centering
\includegraphics[angle=-90,trim=0 0 0 0,clip,width=0.5\textwidth]{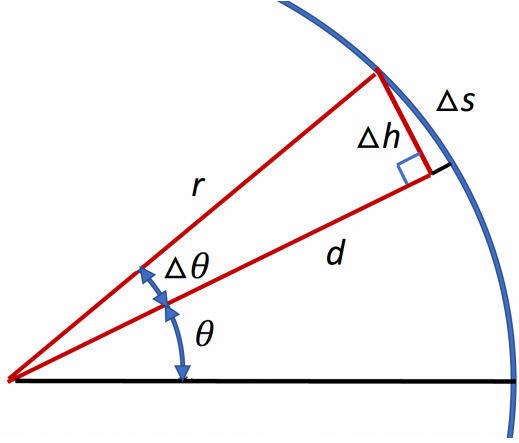}
\caption{The triangle (in red) relevant to Eqs.~(\ref{eq:deriv})-(\ref{eq:lim})
with angle $\Delta \theta$.  The hypotenuse has length $r$, the adjacent
side has length $d$ and the opposite side has length $\Delta h$.  The
arc length of the circle of radius $r$ subtended by $\Delta \theta$ is
$\Delta s$.}
\label{fig:trigill}
\end{figure}

To address this, we calculate the derivative as the defining limit by
writing
\begin{eqnarray}
\frac{\rd}{\rd\theta} \,\sin(\theta) = \lim_{\Delta\theta\rightarrow0}
\frac{\sin(\theta+\Delta\theta) - \sin(\theta)}{\Delta \theta} \, .
\label{eq:deriv}
\end{eqnarray}
From the conventional angle sum identity, based on the geometric
properties of triangles and the notation
in Fig.~\ref{fig:trigill}, we have
\begin{eqnarray}
\sin(\theta+\Delta\theta) &=& \sin(\theta)\cos(\Delta\theta)
+\cos(\theta)\sin(\Delta\theta)
\nonumber\\[5 pt] &=&
\sin(\theta)\,\frac{d}{r} +
\cos(\theta)\,\frac{\Delta h}{r} \, .
\end{eqnarray}
Also from Fig.~\ref{fig:trigill}, it is evident that
\begin{eqnarray}
\lim_{\Delta \theta\rightarrow 0} \frac{d}{r}
&\rightarrow& 1 \, ,
\\[5 pt]
\lim_{\Delta \theta\rightarrow 0} \frac{\Delta h}{r}
&\rightarrow& \frac{\Delta s}{r} \, .
\label{eq:lim}
\end{eqnarray}
Thus,
\begin{eqnarray}
\frac{\rd}{\rd \theta} \, \sin(\theta)
&=& {\cal C}
\cos(\theta) \, ,
\label{eq:sd}
\end{eqnarray}
where
\begin{eqnarray}
 {\cal C}
&=& \lim_{\Delta \theta \rightarrow 0}
\,\frac{\Delta s}{r\,\Delta \theta} = \frac{1}{r}\,\frac{\rd s}{\rd
\theta} \, .
\label{eq:cdef}
\end{eqnarray}
Note that this provides a dimensionally consistent result for the
derivative.  The coefficient ${\cal C}$ can be evaluated by the integration
\begin{eqnarray}
{\cal C}\int_0^\Theta\rd\theta = \frac{1}{r}\int_0^{2\pi r} \rd s \, ,
\end{eqnarray}
where $\Theta$ is the angle of a complete revolution or period of the
sine function
\begin{eqnarray}
\sin(\theta + \Theta) = \sin(\theta) \ ,
\end{eqnarray}
and $2\pi r$ is the circumference of a circle of radius $r$, which yields
\begin{eqnarray}
{\cal C} = \frac{2\pi}{\Theta} \, .
\label{eq:cval}
\end{eqnarray}

Similarly, for the cosine function, we have
\begin{eqnarray}
\frac{\rd}{\rd \theta} \, \cos(\theta)
&=& \lim_{\Delta \theta\rightarrow 0}\,
\frac{\cos(\theta+\Delta \theta)
-\cos(\theta)}{\Delta \theta} \, ,
\end{eqnarray}
where
\begin{eqnarray}
\cos(\theta+\Delta \theta) &=&
\cos(\theta)\cos(\Delta \theta) -
\sin(\theta)\sin(\Delta \theta)
\nonumber\\[5 pt] &=&
\cos(\theta)\,\frac{d}{r} -
\sin(\theta)\,\frac{\Delta h}{r} \, ,
\end{eqnarray}
which gives
\begin{eqnarray}
\frac{\rd}{\rd \theta} \, 
\cos(\theta)
&=& - {\cal C} \sin(\theta) \, .
\label{eq:cderiv}
\end{eqnarray}

\subsection{Exponential function}

The complete exponential function of an angle can be defined in terms of
the complete trigonometric functions.  This generalization recognizes
that the exponential function expressed as Euler's number e raised to a
power is not a complete function. In that case the argument is the
numerical part of the angle expressed in radian units, as shown in the
next section.

The complete-function definition is
\begin{eqnarray}
\exp(\ri\,\theta) &=& \cos(\theta)
+ \ri\,\sin(\theta) \, ,
\label{eq:euf}
\end{eqnarray}
where $\ri^2=-1$.  That is, we do not specify the number raised to a
power.  The derivative follows from the derivatives of the cosine and
sine functions as
\begin{eqnarray}
\frac{\rd}{\rd \theta} \, 
\exp(\ri\,\theta) 
&=& -{\cal C}\sin{\theta} + \ri \, {\cal C} \cos{\theta}
\nonumber\\
&=& \ri\, {\cal C} \exp(\ri\,\theta) \, .
\nonumber\\
&=& \ri\,\frac{2\pi}{\Theta} \, \exp(\ri\,\theta) \, .
\label{eq:deuf}
\end{eqnarray}
We also have $\exp(0)=1$.  With this generalization of the derivative of
the exponential function, one obtains the power series
\begin{eqnarray}
\exp(\ri\,\theta) &=& 
 \sum_{n=0}^\infty \frac{1}{n!}\left(
\frac{2\pi\ri\,\theta}{\Theta}\right)^n \, ,
\label{eq:expeq}
\end{eqnarray}
or
\begin{eqnarray}
\exp(\ri\,\theta) &=& \re^{2\pi\ri\theta/\Theta} \, .
\label{eq:genexp}
\end{eqnarray}
The expressions on the right-hand side of Eqs.~(\ref{eq:expeq}) and
(\ref{eq:genexp}) are well-defined, because the ratio $\theta/\Theta$
is just a number with no unit.

The power series for the complete cosine and sine functions, the real
and imaginary parts of Eq.~(\ref{eq:expeq}), are
\begin{eqnarray}
\cos(\theta) &=& \sum_{n=0}^\infty
\frac{(-1)^n}{(2n)!}
\left(\frac{2\pi\theta}{\Theta}\right)^{2n} \, ,
\label{eq:cosex}
\\
\sin(\theta) &=& \sum_{n=0}^\infty \frac{(-1)^n}{(2n+1)!}
\left(\frac{2\pi\theta}{\Theta}\right)^{2n+1} \, .
\label{eq:sinex}
\end{eqnarray}

If
\begin{eqnarray}
z = \exp(\ri\theta)
\end{eqnarray}
then the complete logarithmic function is
\begin{eqnarray}
\log(z) = \ri\theta + \ri k \Theta  \, ,
\label{eq:clog}
\end{eqnarray}
where $\theta$ is the geometric angle with the dimension of angle and
$k$ is an integer.

To gain some perspective, it is useful to consider differentiation and
integration of the complete logarithmic function to check the
consistency of viewing it as having the dimension of angle.  From
Eq.~(\ref{eq:euf}), we have
\begin{eqnarray}
\frac{\rd z}{\rd \theta} &=& \frac{2\pi\ri}{\Theta}\,z \, ,
\end{eqnarray}
so that
\begin{eqnarray}
\frac{\rd}{\rd z}\log{(z)} = \frac{\rd}{\rd z}\,\ri\,\theta
= \left(\frac{\rd}{\rd \theta}\,\ri\, \theta\right)
\frac{\rd \theta}{\rd z}
= \frac{\Theta}{2\pi}\,\frac{1}{z} \, .
\end{eqnarray}
Thus the derivative of the logarithmic function also has the correct
dimension of angle.  (Recall that $z$ is a dimensionless number.)  To
check the integral of $\log(z)$, we write
\begin{eqnarray}
\frac{\rd}{\rd z}\, z\log{(z)} &=& \log(z) + \frac{\Theta}{2\pi}
= \log(z) + \frac{\rd}{\rd z}\,\frac{\Theta}{2\pi}\,z \, ,
\end{eqnarray}
or
\begin{eqnarray}
\frac{\rd}{\rd z}\left(z\log{(z)}-\frac{\Theta}{2\pi}\,z\right)
&=& \log(z)
\end{eqnarray}
so that
\begin{eqnarray}
\int \rd z \log(z) &=& z \log(z) - \frac{\Theta}{2\pi}\,z + \mbox{
constant} \, ,
\end{eqnarray}
where the constant has the dimension of angle.  Thus the integral of the
logarithmic function also has the proper unit of angle.  Of course, this
expression reduces to the special case commonly used, where the
logarithmic function is assumed to have the base e and the radian is
replaced by 1.  That is, log$\rightarrow$ ln and $\Theta \rightarrow
2\pi$.  In this form the integral of $\ln(z)$ is $z\ln(z) - z$.

The exponential function in Eq.~(\ref{eq:expeq}) may be analytically
continued to real values of the argument to give
\begin{eqnarray}
\exp(\phi) &=& 
 \sum_{n=0}^\infty \frac{1}{n!}\left(
\frac{2\pi\phi}{\Theta}\right)^n \, ,
\label{eq:rexpeq}
\end{eqnarray}
which provides the hyperbolic functions cosh and sinh as
\begin{eqnarray}
\cosh(\phi) &=& \frac{\exp(\phi)+\exp(-\phi)}{2}
=\sum_{n=0}^\infty\frac{1}{(2n)!}
\left(\frac{2\pi\phi}{\Theta}\right)^{2n} \, ,
\\
\sinh(\phi) &=& \frac{\exp(\phi)-\exp(-\phi)}{2}
=\sum_{n=0}^\infty\frac{1}{(2n+1)!}
\left(\frac{2\pi\phi}{\Theta}\right)^{2n+1} \, .
\end{eqnarray}

\subsection{Explicit unit expressions}

\subsubsection{Arbitrary units, $[\theta] =$ {\rm A}}

To connect to the earlier sections in which particular units are
considered, the expressions in the previous section can be given for a
particular, but arbitrary, choice of unit $[\theta] = $ A.  In this case, we employ
the relations $\Theta = \{\Theta\}_{\rm A}$ A and $\theta =
\{\theta\}_{\rm A}$ A to write
\begin{eqnarray}
\frac{\theta}{\Theta} = \frac{\{\theta\}_{\rm A}}{\{\Theta\}_{\rm A}} \, ,
\end{eqnarray}
a ratio that is independent of units. Equation~(\ref{eq:genexp}) may be
written as
\begin{eqnarray}
\exp(\ri\,\theta) &=& b_{\rm A}^{\,\ri\{\theta\}_{\rm A}} \, ,
\end{eqnarray}
where
\begin{eqnarray}
b_{\rm A} = \re^{2\pi/\{\Theta\}_{\rm A}}
\end{eqnarray}
is the base of the exponential function for the unit A.  The
corresponding logarithmic function is
\begin{eqnarray}
\log_{b_{\rm A}}(z) = \ri \{\theta\}_{\rm A} \, ,
\end{eqnarray}
where
\begin{eqnarray}
z = b_{\rm A}^{\,\ri\{\theta\}_{\rm A}} \, . 
\end{eqnarray}
This can be compared to the complete logarithmic function in
Eq.~(\ref{eq:clog}), which has the dimension of angle as its value.

We also have
\begin{eqnarray}
\cos_{\rm A}(\{\theta\}_{\rm A}) &=& \frac{b_{\rm A}^{\,\ri\{\theta\}_{\rm A}} 
+ b_{\rm A}^{\,-\ri\{\theta\}_{\rm A}}}{2} \, , \\
\sin_{\rm A}(\{\theta\}_{\rm A}) &=& \frac{b_{\rm A}^{\,\ri\{\theta\}_{\rm A}} 
- b_{\rm A}^{\,-\ri\{\theta\}_{\rm A}}}{2\,\ri} \, .
\end{eqnarray}

\subsubsection{Radian unit, {\rm A = rad}}

If A is the radian, we have $2\pi/\{\Theta\}_{\rm rad} = 1$,
$b_{\rm rad} = \re$, and
\begin{eqnarray}
\exp(\ri\,\theta) &=& \re^{\ri\{\theta\}_{\rm rad}} \, ,
\label{eq:radexp}
\end{eqnarray}
which is the conventional result.  If
\begin{eqnarray}
z = \re^{\ri\{\theta\}_{\rm rad}} \, ,
\end{eqnarray}
then
\begin{eqnarray}
\log_{\re}(z) = \ln(z) = \ri\{\theta\}_{\rm rad} \, .
\end{eqnarray}

Also
\begin{eqnarray}
\cos_{\rm rad}(\{\theta\}_{\rm rad}) &=& \frac{\re^{\,\ri\{\theta\}_{\rm rad}} 
+ \re^{\,-\ri\{\theta\}_{\rm rad}}}{2} \, , \\
\sin_{\rm rad}(\{\theta\}_{\rm rad}) &=& \frac{\re^{\,\ri\{\theta\}_{\rm rad}} 
- \re^{\,-\ri\{\theta\}_{\rm rad}}}{2\,\ri} \, .
\label{eq:convsin}
\end{eqnarray}

As is evident from Eqs~(\ref{eq:radexp})-(\ref{eq:convsin}), where
${\cal C}=1$, the conventional trigonometric and exponential functions
are implicitly assumed to be based on the radian unit.  For example, the
relation that the cosine function is the derivative of the sine function
with no numerical coefficient assumes that the argument is the numerical
factor of the angle expressed in the radian unit.  Similarly, when the
exponential function is written as the base e raised to the power of the
argument, it is assumed that the argument is the numerical factor of the
angle expressed in the radian unit.

\subsubsection{Revolution or cycle unit, {\rm A = rev}}

In this case, $2\pi/\{\Theta\}_{\rm rev} = 2\pi$, $b_{\rm rev} =
\re^{2\pi}$, and
\begin{eqnarray}
\exp(\ri\,\theta) &=& \re^{2\pi\ri\{\theta\}_{\rm rev}} \, .
\end{eqnarray}

It is important to be aware of the units being considered, because it is
also often implicitly assumed that the revolution ({\it i.e.,} cycle or
period), rather than the radian, is the suppressed unit, particularly
where the angle denotes phase.  For example, wavelength $\lambda$ is
often taken to be the distance over which a spatially periodic function
undergoes a phase change corresponding to one revolution or cycle rather
than one radian.  In the latter case, it is often called the reduced
wavelength $\lbar$.  The risk of not taking this into account can
and does lead to errors of $2\pi$.  Similarly, frequencies, are commonly
expressed in Hz or cycles/s rather than radians/s, which can also lead
to the same error.

\subsubsection{Dimensionless functions}

Various forms of the exponential and logarithmic functions are described
above in terms of units for the dimension of angle or phase.  For purely
numerical (dimensionless) applications, the radian-based forms are
commonly used with no reference to angle units.  In such applications,
we have
\begin{eqnarray}
y = \re^x\, ,
\end{eqnarray}
and
\begin{eqnarray}
x = \ln(y) \, ,
\end{eqnarray}
where $x$ and $y$ are dimensionless, possibly complex, numbers.

Another commonly used dimensionless base is 10, where
\begin{eqnarray}
y = 10^{\,x} \, ,
\end{eqnarray}
and
\begin{eqnarray}
x = \log_{10}(y) \, .
\end{eqnarray}
In either of these applications, there is no reference to angle or
phase.

\section{Applications}

The use of complete transcendental functions can resolve some apparent
mismatches of angle units that appear in commonly used equations in
physics.  To examine this, it is useful to distinguish between ``unit
analysis'' and ``dimensional analysis''.  As considered here,
dimensional analysis is more general in the sense that a geometrical
angle or phase angle may be expressed in various units, such as degrees,
radians, and revolutions or cycles, but in all cases, it has the
dimension of angle, which we denote by \textsf{A}.  This can be
expressed by writing $<\!\theta\!>\, = \textsf{A}$, where $\theta$ is an
angle and the brackets are used to denote dimension, in contrast to the
notation for the unit $[\theta] =$ rad, for example.  Other dimensions
that we will consider here are time, length, mass, and charge denoted by
\textsf{T}, \textsf{L}, \textsf{M}, and \textsf{Q}, respectively.
(While the ampere, with dimension ``current'' is one of the traditional
base units of the SI, the definition of the ampere involves the defining
charge of the electron,  Hence, it is natural, when considering electric
units, to focus on the unit couloumb and its associated dimension,
electrical charge.) In all cases, the dimensions on either side of an
equation must be the same.

In the following, examples are given where a dimensional analysis
based on complete functions resolves an apparent angle unit ambiguity.

\subsection{Centripetal acceleration}

A familiar relation in physics is the equation for centripetal
acceleration $a_{\rm c}$ of a mass, in uniform circular motion,
conventionally given by
\begin{eqnarray}
a_{\rm c} = \frac{v^2}{r} = r\omega^2\, ,
\label{eq:cacc}
\end{eqnarray}
where $v$ is the linear velocity of the mass, $r$ is the radius of the
path, and $\omega$ is the rotational frequency of the mass.  The SI
units of the two terms on the right-hand end of that equation are m
s$^{-2}$ and m rad$^2$ s$^{-2}$.  There is an obvious mismatch between
these two terms, which is also reflected in the dimensional analysis
which gives \textsf{L} \textsf{T}$^{-2}$ and \textsf{L} \textsf{A}$^2$
\textsf{T}$^{-2}$ for the latter two terms, whereas $a_{\rm c}$ has the
dimension \textsf{L} \textsf{T}$^{-2}$.   This difference can be
resolved by deriving the complete expression for the acceleration from
basic principles.

Consider the uniform circular motion of a
mass described by its position in cylindrical coordinates 
$\bm r(r,\phi)$.  For motion in
the $\bm{\hat i},~\bm{\hat j}$ plane, we have
\begin{eqnarray}
\bm r(r,\phi) &=& 
r \cos(\phi) \, \bm{\hat i} + r \sin(\phi) \, \bm{\hat j} \, ,
\end{eqnarray}
where $\phi = \omega t$.
The centripetal acceleration is
\begin{eqnarray}
\bm a_{\rm c} = \frac{\rd^2}{\rd t^2}\, \bm
r(r,\phi) 
= \omega^2 \, \frac{\rd^2}{\rd \phi^2}\,
\bm r(r,\phi)  \, .
\end{eqnarray}
From Eqs.~(\ref{eq:sd}), (\ref{eq:cval}), and (\ref{eq:cderiv}), we have
\begin{eqnarray}
\frac{\rd^2}{\rd \phi^2}\, \cos(\phi) &=& 
- \left(\frac{2\pi}{\Theta}\right)^2
\cos(\phi) \, ,
\\[5 pt]
\frac{\rd^2}{\rd \phi^2}\, \sin(\phi) &=& 
- \left(\frac{2\pi}{\Theta}\right)^2
\sin(\phi) \, ,
\end{eqnarray}
so that
\begin{eqnarray}
\bm a_{\rm c} &=& - \left(\frac{2\pi\omega}{\Theta}\right)^2
\bm r(r,\phi)  \, ,
\end{eqnarray}
or
\begin{eqnarray}
a_{\rm c} = r \left(\frac{2\pi\omega}{\Theta}\right)^2
\label{eq:ccacc}
\end{eqnarray}
This has proper dimensionality of \textsf{L T}$^{-2}$, which resolves
the disagreement in Eq.~(\ref{eq:cacc}).

For the radian as the angle unit, we have $\Theta = 2\pi$ rad and
\begin{eqnarray}
a_{\rm c} &=& 
 r \omega_{\rm rad}^2 \, ,
\end{eqnarray}
where $\omega_{\rm rad} \equiv \omega/(1~{\rm rad})$.  This means that
$a_{\rm c} = r\omega^2$ is not a complete equation, although $a_{\rm c}
= v^2/r$ and Eq.~(\ref{eq:ccacc}) are, which resolves the
mismatch.\cite{error}

\subsection{Volume integration}

If we do a transformation of a Cartesian volume element to a
volume element in spherical coordinates, we have
\begin{eqnarray}
<\! \rd x \,\rd y \,\rd z \!>\, &=& \textsf{L}^3 \, ,
\\
<\! r^2 \,\rd r \, \sin{\theta} \, \rd\theta \,\rd\phi \!>\, &=& \textsf{L}^3
~\textsf{A}^2 \, .
\end{eqnarray}
How can this be made consistent?

The transformed coordinates are
\begin{eqnarray}
x &=& r \sin{(\theta)}\cos{(\phi)} \, , \\[5 pt]
y &=& r \sin{(\theta)}\sin{(\phi)} \, , \\[5 pt]
z &=& r \cos{(\theta)} \, .
\end{eqnarray}
The transformation is given by the Jacobian determinant
\begin{eqnarray}
&&\rd x \, \rd y \, \rd z = \left|\begin{array}{c@{\quad}c@{\quad}c}
\tfrac{\partial x}{\partial r} & \tfrac{\partial x}{\partial \theta}
& \tfrac{\partial x}{\partial \phi} \\[10 pt]
\tfrac{\partial y}{\partial r} & \tfrac{\partial y}{\partial \theta}
& \tfrac{\partial y}{\partial \phi} \\[10 pt]
\tfrac{\partial z}{\partial r} & \tfrac{\partial z}{\partial \theta}
& \tfrac{\partial z}{\partial \phi} \\
\end{array}\right|
\rd r \, \rd \theta \, \rd \phi
\nonumber\\[10 pt] 
&&\qquad= \left(\frac{2\pi}{\Theta}\right)^2 
\left|\begin{array}{c@{\quad}c@{\quad}c}
\sin{(\theta)}\cos{(\phi)} & 
r\cos{(\theta)}\cos{(\phi)}
&-r\sin{(\theta)}\sin{(\phi)} \\
\sin{(\theta)}\sin{(\phi)} & 
r\cos{(\theta)}\sin{(\phi)}
&r\sin{(\theta)}\cos{(\phi)} \\
\cos{(\theta)} & -r\sin{(\theta)} & 0 \\
\end{array}\right| 
\rd r \, \rd \theta \, \rd \phi \qquad
\nonumber\\[10 pt] 
&&\qquad= \left(\frac{2\pi}{\Theta}\right)^2 \,
r^2 \,\rd r \, 
\sin{(\theta)}
\rd \theta \, \rd \phi \, ,
\end{eqnarray}
which has the dimension \textsf{L}$^3$, as required.
Thus, for an integral, we have
\begin{eqnarray}
&&\int_{-\infty}^{\infty}\rd x \int_{-\infty}^{\infty}\rd y
\int_{-\infty}^{\infty}\rd z \, f(x,y,z)
= \left(\frac{2\pi}{\Theta}\right)^2 \int_0^{\infty}\rd r \, r^2
\int_{-\Theta/2}^{\Theta/2}\rd\theta\sin{(\theta)}
\int_0^\Theta\rd \phi
\nonumber\\[5 pt]
&&\qquad\qquad\qquad\qquad\qquad\times
f\big(r\sin{(\theta)}\cos{(\phi)},
r\sin{(\theta)}\sin{(\phi)},
r\cos{(\theta)}\big) \, .
\end{eqnarray} 
If the unit for angles is the radian, then $\Theta = 2\pi$ rad, and
\begin{eqnarray}
\sin(\theta) &=& \sin_{\rm rad}(\{\theta\}_{\rm rad}) 
\equiv \sin_{\rm rad}(\theta_{\rm rad}) \qquad etc. \\
\frac{2\pi}{\Theta}\,\rd \theta &=& \rd\{\theta\}_{\rm rad} 
\equiv \rd\theta_{\rm rad} \qquad etc.
\end{eqnarray}
This gives
\begin{eqnarray}
&&\int_{-\infty}^{\infty}\rd x \int_{-\infty}^{\infty}\rd y
\int_{-\infty}^{\infty}\rd z \, f(x,y,z)
= \int_0^{\infty}\rd r \, r^2
\int_{-\pi}^\pi\rd\theta_{\rm rad}\sin_{\rm rad}{(\theta_{\rm rad})}
\int_0^{2\pi}\rd \phi_{\rm rad}
\nonumber\\[5 pt]
&&\qquad\qquad\times
f\big(r\sin_{\rm rad}{(\theta_{\rm rad})}\cos_{\rm rad}{(\phi_{\rm rad})},
r\sin_{\rm rad}{(\theta_{\rm rad})}\sin_{\rm rad}{(\phi_{\rm rad})},
r\cos_{\rm rad}{(\theta_{\rm rad})}\big) \, ,
\end{eqnarray} 
which reduces to the conventional result if the radian label is dropped:
\begin{eqnarray}
&&\int_{-\infty}^{\infty}\rd x \int_{-\infty}^{\infty}\rd y
\int_{-\infty}^{\infty}\rd z \, f(x,y,z)
= \int_0^{\infty}\rd r \, r^2
\int_{-\pi}^\pi\rd\theta\sin{(\theta)}
\int_0^{2\pi}\rd \phi
\nonumber\\[5 pt]
&&\qquad\qquad\times
f\big(r\sin{(\theta)}\cos{(\phi)},
r\sin(\theta)\sin{(\phi)},
r\cos{(\theta)}\big) \, .
\end{eqnarray} 

\subsection{Water waves}

The problem of the phase velocity of shallow water waves provides an
interesting case about unit compatibility.  

The explicit conventional expression for the phase velocity is
\begin{eqnarray}
c_{\rm p} = \frac{\omega}{k} = \frac{\sqrt{gk\tanh{(kh)}}}{k}
=\sqrt{gh}\,\sqrt{\frac{\tanh{(kh)}}{kh}} 
\, ,
\end{eqnarray}
corresponding to 
\begin{eqnarray}
\omega^2 = gk\tanh(kh) \, ,
\label{eq:water}
\end{eqnarray}
where $\omega$ is the frequency of the wave, $k=1/\lbar$ is the wavenumber, $g$
is the acceleration of gravity, and $h$ is the depth of the water.
The corresponding dimensions are $<\!\omega^2\!>\, =
\textsf{A}^2~\textsf{T}^{-2}$, $<\!k\!>\, = \textsf{A}~\textsf{L}^{-1}$,
$<\!g\!>\, = \textsf{L~T}^{-2}$, and $<\!h\!>\, = \textsf{L}$.

The problem is that the left-hand side of Eq.~(\ref{eq:water}) has the
dimension \textsf{A$^2$ T$^{-2}$} and the right-hand side has dimension
\textsf{A T$^{-2}$}, so that one angle dimension appears to have been
lost in the derivation.  This contradiction can be fixed by repeating
the derivation with angles properly taken into account.
Eq.~(\ref{eq:water}) follows from solving the equation for the velocity
potential given by\cite{1978040}
\begin{eqnarray}
\frac{\partial^2\phi}{\partial t^2} + g\,\frac{\partial\phi}{\partial z}
= 0  \qquad \mbox{at } z = 0\, ,
\end{eqnarray}
with the boundary condition
\begin{eqnarray}
\left.\frac{\partial\phi}{\partial z}\right|_{z = -h} = 0 \, .
\end{eqnarray}
The solution is 
\begin{eqnarray}
\phi = \frac{\cosh k(z+h)}{k\sinh kh}\,\omega a \sin(\bm k\cdot\bm x -
\omega t) \, ,
\end{eqnarray}
where $a$ is a normalization constant and $k_z = 0$.  Differentiation
gives
\begin{eqnarray}
\frac{\partial^2\phi}{\partial t^2} =
-\left(\frac{2\pi\omega}{\Theta}\right)^2\phi \, ,
\end{eqnarray}
so that
\begin{eqnarray}
\omega^2
= \frac{g}{\phi}\,\left(\frac{\Theta}{2\pi}\right)^2
\left.\frac{\partial\phi}{\partial z}
\right|_{z=0} =  gk\,\frac{\Theta}{2\pi}\,\tanh kh \, .
\end{eqnarray}
The right-hand side has the correct dimensions of \textsf{A$^2$
T$^{-2}$}.

\subsection{Units for the cyclotron resonance frequency}
\label{sec:cyclotron}

The relevant conventional equation is
\begin{eqnarray}
\omega = \frac{qB_{\rm c}}{m} \, ,
\label{eq:crf}
\end{eqnarray}
where $\omega$ is the frequency, $q/m$ is the charge to mass ratio of
the particle, and $B_{\rm c}$ is the classical magnetic field.  The
dimensions are $<\!\omega\!>\, = \textsf{A~T}^{-1}$, $<\!q/m\!>\, =
\textsf{Q~M}^{-1}$, and $<\!B_{\rm c}\!>\, = \textsf{M
T}^{-1}~\textsf{Q}^{-1}$.

The left-hand side of Eq.~(\ref{eq:crf}) has dimension \textsf{A
T$^{-1}$} and the right-hand side works out to \textsf{T$^{-1}$}, so
there is a mismatch of the angle dimension and the equation is not
complete.  [Moreover, because the right-hand side has dimension
\textsf{T$^{-1}$}, the current SI prescribes that the unit is s$^{-1}$,
or Hz, which is not correct [see Eq.~(\ref{eq:cyfr})].\cite{2019SI}  As
a result, \emph{Mathematica}\cite{math}, which provides calculations
with units, gets regular bug reports about this.  To resolve this
problem, we work out the complete version of this equation.]

For motion of a particle of mass $m$ and charge $q$ in the $\bm{\hat
i},~\bm{\hat j}$ plane along a circular path of radius r, centered at
$\bm x = 0$, we have 
\begin{eqnarray}
\bm x &=& r \, \cos(\omega t) ~ \bm{\hat i} + r \, \sin
(\omega t) ~ \bm{\hat j}
\\[5 pt]
\bm v =
\frac{\rm d}{{\rm d} t} \, \bm x &=&
\frac{\rd}{\rd t} 
\left[r \, \cos(\omega t) ~ \bm{\hat i} + r \, \sin
(\omega t) ~ \bm{\hat j}\right]
\\[5 pt]
&=& \frac{2\pi\omega}{\Theta} \left[-r \, \sin(\omega t) ~ \bm{\hat i} 
+ r \, \cos (\omega t) ~ \bm{\hat j}\right]
\\[5 pt]
\bm a = \frac{\rd}{\rd t} \, \bm v
&=& -\left(\frac{2\pi\omega}{\Theta}\right)^2 
\left[r \, \cos(\omega t) ~ \bm{\hat i} 
+ r \, \sin(\omega t) ~ \bm{\hat j}\right]
\\[5 pt]
&=& -\left(\frac{2\pi\omega}{\Theta}\right)^2 \bm x
\end{eqnarray}
For a magnetic field given by $\bm B_{\rm c} = - B_{\rm c}
\bm{\hat k}$, the classical force is
\begin{eqnarray}
\bm F_{\rm c} &=& q \bm v \times \bm B_{\rm c} = -q
 B_{\rm c}\,\frac{2\pi\omega}{\Theta}
 \left[r \, \sin(\omega t) ~ \bm{\hat j} 
+ r \, \cos (\omega t) ~ \bm{\hat i}\,\right]
\\[5 pt]
&=& -q B_{\rm c} \frac{2\pi\omega}{\Theta} \bm x
\end{eqnarray}
Thus from Newton's law, $\bm F_{\rm c} = m\bm a$, we have
\begin{eqnarray}
q B_{\rm c} = m \, \frac{2\pi\omega}{\Theta}
\end{eqnarray}
or
\begin{eqnarray}
\omega = \frac{qB_{\rm c}}{m}\,\frac{\Theta}{2\pi}
\end{eqnarray}
This is the complete equation, and the dimensions, \textsf{A T}$^{-1}$,
match.

In SI units, the frequency is
\begin{eqnarray}
\omega &=& \left\{\frac{qB_{\rm c}}{m}\right\}\times
\left\{\begin{array}{l}\mbox{rad~s}^{-1} \\
\frac{\mbox{cycles~s}^{-1}}{\textstyle 2\pi} 
= \frac{\mbox{Hz}}{\textstyle 2\pi}
\end{array} \right.  \, .
\label{eq:cyfr}
\end{eqnarray}

\subsection{Classical pendulum}
\label{sec:cp}

\begin{figure}[t]
\includegraphics[trim=0 15 0 15,clip,width=0.5\textwidth]{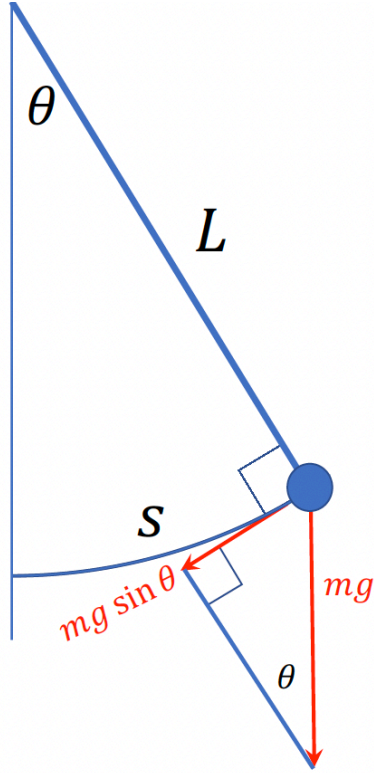}
\caption{Classical pendulum}
\label{fig:pendulum}
\end{figure}

The conventional expression for the small-oscillation frequency $\omega$
of a classical pendulum is
\begin{eqnarray}
\omega = \sqrt{\frac{g}{L}} \, ,
\end{eqnarray}
where $g$ is the acceleration of gravity and $L$ is the length of the
pendulum arm.  The current SI unit for $\sqrt{g/L}$ is s$^{-1}$, which
suggests the frequency is in Hz, but, as for the case of the cyclotron
frequency above, this is not correct, as shown by considering the
complete equation calculation.

The equation of motion is
\begin{eqnarray}
m\,\frac{\rd^2 s(t)}{\rd t^2} &=& - mg \sin(\theta(t))\, ,
\end{eqnarray}
where $\theta$ and $s$ are the angle and arc length from the lowest
position of the pendulum, as indicated in Fig.~\ref{fig:pendulum}.  The
mass cancels out of the equation.  We consider small displacements from
the lowest point, so that
\begin{eqnarray}
\sin\big(\theta(t)\big) = \frac{2\pi\theta(t)}{\Theta} + \dots \, ,
\end{eqnarray}
according to Eq.~(\ref{eq:sinex}).  From
Eqs.~(\ref{eq:cdef}) and (\ref{eq:cval}), we have
\begin{eqnarray}
\frac{2\pi}{\Theta}\int\rd\theta = \frac{1}{L}\int \rd s \, ,
\end{eqnarray}
which provides the relation
\begin{eqnarray}
s(t) = \frac{2\pi L}{\Theta} \, \theta(t) \, .
\end{eqnarray}
We thus have
\begin{eqnarray}
\frac{\rd^2\theta(t)}{\rd t^2} + \frac{g}{L} \, \theta(t) = 0\, .
\end{eqnarray}
The general solution is
\begin{eqnarray}
\theta(t) = a\sin(\omega t) + b\cos(\omega t) \, ,
\end{eqnarray}
and so the complete second derivative
\begin{eqnarray}
\frac{\rd^2\theta(t)}{\rd t^2} &=& - \left(\frac{2\pi\omega}{\Theta}\right)^2
\theta(t)
\end{eqnarray}
gives
\begin{eqnarray}
\omega = \sqrt{\frac{g}{L}}\,\frac{\Theta}{2\pi} \, ,
\end{eqnarray}
which has the dimension \textsf{A T}$^{-1}$.
Finally, the proper SI expression for $\omega$ is
\begin{eqnarray}
\omega &=& \sqrt{\left\{\frac{g}{L}\right\}} \mbox{ rad s}^{-1}  \, .
\label{eq:pendfr}
\end{eqnarray}

\subsection{Jacobi elliptic functions}

The Jacobi elliptic functions ${\rm sn}(u,k)$ and ${\rm cn}(u,k)$ may be
viewed as generalizations of the sine and cosine functions because
\cite{wolf}
\begin{eqnarray}
{\rm sn}(u,0) &=& \sin{u}  \, ,
\\[5 pt]
{\rm cn}(u,0) &=& \cos{u} \, .
\end{eqnarray}
It is natural to seek the generalization of the 
differentiation formula in Eq.~(\ref{eq:sd}).

The functions, together with ${\rm dn}(u,k)$, are defined by
\begin{eqnarray}
u &=& \int_0^\phi \, \frac{\rd
\phi^\prime}{\sqrt{1-k^2\sin^2{\phi^\prime}}} \, ,
\label{eq:udef}
\end{eqnarray}
\begin{eqnarray}
{\rm sn}(u,k) &=& \sin{\phi} \, ,
\\[10 pt]
{\rm cn}(u,k) &=& \cos{\phi} \, ,
\\[10 pt]
{\rm dn}(u,k) &=& \sqrt{1-k^2\sin^2{\phi}} \, .
\end{eqnarray}
From Eq.~(\ref{eq:udef}), we have
\begin{eqnarray}
\frac{\rd u}{\rd \phi} &=& \frac{1}{\sqrt{1-k^2\sin^2{\phi}}} \, ,
\end{eqnarray}
and so
\begin{eqnarray}
\frac{\rd}{\rd u} \, {\rm sn}(u,k) &=& 
\frac{\rd}{\rd u} \, \sin{\phi} =
\left(\frac{\rd}{\rd \phi}\, \sin{\phi}\right)\frac{\rd \phi}{\rd u}
\nonumber\\[10 pt]
&=& \frac{2\pi}{\Theta}\,\cos{\phi}\, \sqrt{1-k^2\sin^2{\phi}}
\nonumber\\[10 pt]
&=& \frac{2\pi}{\Theta}\,{\rm cn}(u,k)\,{\rm dn}(u,k) \, ,
\end{eqnarray}
which reduces to the conventional result for the derivative of ${\rm
sn}(u,k)$ if rad $\rightarrow 1$ and agrees with Eq.~(\ref{eq:sd}) if
$k=0$.

\section{Conclusion}

Here, we review some key points relevant to the treatment of angles and
their units in equations of physics.  In this context, the term angle is
used to mean both plane angle and phase angle.

A guiding principle is that the use and properties of units ({\it e.g.}
their dimensions) is a matter of choice.  Units are not a property of
nature, but rather they provide a rational method for quantitatively
describing natural phenomena with mathematical equations.  Of course,
the choice is not completely arbitrary, because the units must fit into
a logically consistent framework.

We may choose to assign to angles an independent dimension with
associated units, or not.  The current practice is generally to treat
angles simply as numbers and view them as dimensionless.  However, as
should be evident from the forgoing sections, this point of view makes
unstated assumptions.  In particular, it assumes that the number
representing an angle is the numerical factor of the angle expressed in
the radian unit.  This view also assumes that the radian unit is not
actually a unit but is just the number one.  An advantage of this
approach is that it represents the status quo and as such requires no
further thought or analysis.  However, this leads to ambiguities and
errors as described in the forgoing examples.  Because the radian is a
coherent unit in the SI, there is no numerical factor when the radian is
restored, as illustrated by Eqs.~(\ref{eq:cyfr}) and (\ref{eq:pendfr}).
On the other hand, Hz is not a coherent unit as seen in
Eq.~(\ref{eq:cyfr}) where there is an additional factor of $1/2\pi$.

Such ambiguities and errors can be avoided by assigning an independent
dimension to angles and treating this dimension in a consisent way,
insuring the correct dimension for the equations of physics, as well as
insuring the proper units among the various units that might be used
for angles.  The use of complete functions and complete equations is a
powerful tool to insure a consistent treatment of dimension and units.
An important advantage is that it eliminates errors of a factor of
$2\pi$ as shown in the examples.  The assignment of a dimension to
angles also raises awareness that there are built-in hidden assumptions
in the widely accepted way in which angles and trigonometric functions
of angles are treated.  Moreover, the analysis of the consequences of
assigning a dimension to angles provides a generalization of the
relevant equations, so that they become complete equations.  As already
noted, complete equations do not assume that the quantities appearing in
them are expressed in particular units.

To elaborate on this point, we consider the example, often used by
theoretical physicists, of setting the speed of light $c$ to 1.
\cite{1998165}  
When this is done, $c$ disappears from equations that
would otherwise show it.  Then we have $E=m$ instead of $E=mc^2$, which
introduces an ambiguity between energy and mass.  Restoration of $c$ at
the end of a calculation requires judgement by the practitioner.  This
is completely analogous to the practice of setting ${\cal C} =
2\pi/\Theta$, as defined in the preceding sections, equal to 1, which is
the default for the ``radian assumption''.  If ${\cal C}$ is restored,
it shows up as an additional parameter in some equations, just as $c$
shows up when it is not 1.

The cost of this restoration of ${\cal C}=2\pi/\Theta$ is the task of
determining where it belongs in equations in which it has been replaced
by 1, by default.  In this paper, the generalization has been worked out
for transcendental functions and applied in various examples.

\bibliography{refs,misc}

\begin{thebibliography}{60}%
\makeatletter
\providecommand \@ifxundefined [1]{%
 \@ifx{#1\undefined}
}%
\providecommand \@ifnum [1]{%
 \ifnum #1\expandafter \@firstoftwo
 \else \expandafter \@secondoftwo
 \fi
}%
\providecommand \@ifx [1]{%
 \ifx #1\expandafter \@firstoftwo
 \else \expandafter \@secondoftwo
 \fi
}%
\providecommand \natexlab [1]{#1}%
\providecommand \enquote  [1]{``#1''}%
\providecommand \bibnamefont  [1]{#1}%
\providecommand \bibfnamefont [1]{#1}%
\providecommand \citenamefont [1]{#1}%
\providecommand \href@noop [0]{\@secondoftwo}%
\providecommand \href [0]{\begingroup \@sanitize@url \@href}%
\providecommand \@href[1]{\@@startlink{#1}\@@href}%
\providecommand \@@href[1]{\endgroup#1\@@endlink}%
\providecommand \@sanitize@url [0]{\catcode `\\12\catcode `\$12\catcode
  `\&12\catcode `\#12\catcode `\^12\catcode `\_12\catcode `\%12\relax}%
\providecommand \@@startlink[1]{}%
\providecommand \@@endlink[0]{}%
\providecommand \url  [0]{\begingroup\@sanitize@url \@url }%
\providecommand \@url [1]{\endgroup\@href {#1}{\urlprefix }}%
\providecommand \urlprefix  [0]{URL }%
\providecommand \Eprint [0]{\href }%
\providecommand \doibase [0]{http://dx.doi.org/}%
\providecommand \selectlanguage [0]{\@gobble}%
\providecommand \bibinfo  [0]{\@secondoftwo}%
\providecommand \bibfield  [0]{\@secondoftwo}%
\providecommand \translation [1]{[#1]}%
\providecommand \BibitemOpen [0]{}%
\providecommand \bibitemStop [0]{}%
\providecommand \bibitemNoStop [0]{.\EOS\space}%
\providecommand \EOS [0]{\spacefactor3000\relax}%
\providecommand \BibitemShut  [1]{\csname bibitem#1\endcsname}%
\let\auto@bib@innerbib\@empty
\bibitem [{\citenamefont {Quincey}\ \emph {et~al.}(2019)\citenamefont
  {Quincey}, \citenamefont {Mohr},\ and\ \citenamefont {Phillips}}]{2019063}%
  \BibitemOpen
  \bibfield  {author} {\bibinfo {author} {\bibfnamefont {Paul}\ \bibnamefont
  {Quincey}}, \bibinfo {author} {\bibfnamefont {Peter~J.}\ \bibnamefont
  {Mohr}}, \ and\ \bibinfo {author} {\bibfnamefont {William~D.}\ \bibnamefont
  {Phillips}},\ }\bibfield  {title} {\enquote {\bibinfo {title} {Angles are
  inherently neither length ratios nor dimensionless},}\ }\href@noop {}
  {\bibfield  {journal} {\bibinfo  {journal} {Metrologia}\ }\textbf {\bibinfo
  {volume} {56}},\ \bibinfo {eid} {043001} (\bibinfo {year}
  {2019})}\BibitemShut {NoStop}%
\bibitem [{\citenamefont {Baptiste}\ and\ \citenamefont
  {Fourier}(1822)}]{1822001}%
  \BibitemOpen
  \bibfield  {author} {\bibinfo {author} 
  {\bibfnamefont {Jean-Baptiste-Joseph}}\
  \bibnamefont {Fourier, }}\href@noop {} {\emph {\bibinfo {title} {Th\'eorie
  analytique de la chaleur}}}\ (\bibinfo  {publisher} {Chez Firmin Didot,
  P\`ere et fils},\ \bibinfo {address} {Paris},\ \bibinfo {year}
  {1822}), see Sec.~160 {\it et seq}\BibitemShut {NoStop}%
\bibitem [{\citenamefont {Maxwell}(1870)}]{1870001}%
  \BibitemOpen
  \bibfield  {author} {\bibinfo {author} {\bibfnamefont {James~C.}\
  \bibnamefont {Maxwell}},\ }\bibfield  {title} {\enquote {\bibinfo {title}
  {Mathematics and physics},}\ }in\ \href@noop {} {\emph {\bibinfo {booktitle}
  {Report of the Fortieth Meeting of the British Association for the
  Advancement of Science}}},\ \bibinfo {editor} {edited by\ \bibinfo {editor}
  {\bibfnamefont {John}\ \bibnamefont {Murray}}}\ (\bibinfo  {publisher}
  {London},\ \bibinfo {address} {Liverpool},\ \bibinfo {year} {1870})\ pp.\
  \bibinfo {pages} {1--9}\BibitemShut {NoStop}%
\bibitem [{\citenamefont {Maxwell}(1873)}]{1873001}%
  \BibitemOpen
  \bibfield  {author} {\bibinfo {author} {\bibfnamefont {James~Clerk}\
  \bibnamefont {Maxwell}},\ }\href@noop {} {\emph {\bibinfo {title} {A Treatise
  On Electricity And Magnetism}}},\ Vol.~\bibinfo {volume} {1}\ (\bibinfo
  {publisher} {Macmillan and Co.},\ \bibinfo {address} {London},\ \bibinfo
  {year} {1873})\BibitemShut {NoStop}%
\bibitem [{\citenamefont {Muir}(1910)}]{1910001}%
  \BibitemOpen
  \bibfield  {author} {\bibinfo {author} {\bibfnamefont {T.}~\bibnamefont
  {Muir}},\ }\bibfield  {title} {\enquote {\bibinfo {title} {The term `radian'
  in trigonometry},}\ }\href@noop {} {\bibfield  {journal} {\bibinfo  {journal}
  {Nature}\ }\textbf {\bibinfo {volume} {83}},\ \bibinfo {pages} {459--460}
  (\bibinfo {year} {1910})}\BibitemShut {NoStop}%
\bibitem [{\citenamefont {Brinsmade}(1936)}]{1936002}%
  \BibitemOpen
  \bibfield  {author} {\bibinfo {author} {\bibfnamefont {J.~B.}\ \bibnamefont
  {Brinsmade}},\ }\bibfield  {title} {\enquote {\bibinfo {title} {Plane and
  solid angles their pedagogic value when introduced explicitly},}\ }\href@noop
  {} {\bibfield  {journal} {\bibinfo  {journal} {Phys. Teach.}\ }\textbf
  {\bibinfo {volume} {4}},\ \bibinfo {pages} {175--179} (\bibinfo {year}
  {1936})}\BibitemShut {NoStop}%
\bibitem [{\citenamefont {Page}(1961)}]{1961013}%
  \BibitemOpen
  \bibfield  {author} {\bibinfo {author} {\bibfnamefont {Chester~H.}\
  \bibnamefont {Page}},\ }\bibfield  {title} {\enquote {\bibinfo {title}
  {Physical entities and mathematical representation},}\ }\href@noop {}
  {\bibfield  {journal} {\bibinfo  {journal} {J. Res. Natl. Bur. Stand. Sect.
  B}\ }\textbf {\bibinfo {volume} {65B}},\ \bibinfo {pages} {227--235}
  (\bibinfo {year} {1961})}\BibitemShut {NoStop}%
\bibitem [{\citenamefont {Romain}(1962)}]{1962014}%
  \BibitemOpen
  \bibfield  {author} {\bibinfo {author} {\bibfnamefont {Jacques~E.}\
  \bibnamefont {Romain}},\ }\bibfield  {title} {\enquote {\bibinfo {title}
  {Angle as a fourth fundamental quantity},}\ }\href@noop {} {\bibfield
  {journal} {\bibinfo  {journal} {J. Res. Natl. Bur. Stand. Sect. B}\ }\textbf
  {\bibinfo {volume} {66B}},\ \bibinfo {pages} {97--100} (\bibinfo {year}
  {1962})}\BibitemShut {NoStop}%
\bibitem [{\citenamefont {Page}(1978)}]{1978038}%
  \BibitemOpen
  \bibfield  {author} {\bibinfo {author} {\bibfnamefont {Chester~H.}\
  \bibnamefont {Page}},\ }\bibfield  {title} {\enquote {\bibinfo {title}
  {Classes of units in the SI},}\ }\href@noop {} {\bibfield  {journal}
  {\bibinfo  {journal} {Am. J. Phys.}\ }\textbf {\bibinfo {volume} {46}},\
  \bibinfo {pages} {78--79} (\bibinfo {year} {1978})}\BibitemShut {NoStop}%
\bibitem [{\citenamefont {Stiehler}(1978)}]{1978039}%
  \BibitemOpen
  \bibfield  {author} {\bibinfo {author} {\bibfnamefont {D.}~\bibnamefont
  {Stiehler}},\ }\bibfield  {title} {\enquote {\bibinfo {title} {Getting the
  right angle},}\ }\href@noop {} {\bibfield  {journal} {\bibinfo  {journal}
  {ASEE}\ }\textbf {\bibinfo {volume} {5}},\ \bibinfo {pages} {3} (\bibinfo
  {year} {1978})}\BibitemShut {NoStop}%
\bibitem [{\citenamefont {Phillips}(1978)}]{1978040}%
  \BibitemOpen
  \bibfield  {author} {\bibinfo {author} {\bibfnamefont {O.~M.}\ \bibnamefont
  {Phillips}},\ }\href@noop {} {\emph {\bibinfo {title} {The Dynamics of the
  Upper Ocean}}},\ \bibinfo {edition} {2nd}\ ed.\ (\bibinfo  {publisher}
  {Cambridge University Press},\ \bibinfo {address} {Cambridge, United
  Kingdom},\ \bibinfo {year} {1978})\BibitemShut {NoStop}%
\bibitem [{\citenamefont {Boer}(1979)}]{1979035}%
  \BibitemOpen
  \bibfield  {author} {\bibinfo {author} {\bibfnamefont {Jan~de}\ \bibnamefont
  {Boer}},\ }\bibfield  {title} {\enquote {\bibinfo {title} {Group properties
  of quantities and units},}\ }\href@noop {} {\bibfield  {journal} {\bibinfo
  {journal} {Am. J. Phys.}\ }\textbf {\bibinfo {volume} {47}},\ \bibinfo
  {pages} {818--819} (\bibinfo {year} {1979})}\BibitemShut {NoStop}%
\bibitem [{\citenamefont {Page}(1979)}]{1979036}%
  \BibitemOpen
  \bibfield  {author} {\bibinfo {author} {\bibfnamefont {Chester}\ \bibnamefont
  {Page}},\ }\bibfield  {title} {\enquote {\bibinfo {title} {Rebuttal to de
  Boer's `group properties of quantities and unit'},}\ }\href@noop {}
  {\bibfield  {journal} {\bibinfo  {journal} {Am. J. Phys.}\ }\textbf {\bibinfo
  {volume} {47}},\ \bibinfo {pages} {820} (\bibinfo {year} {1979})}\BibitemShut
  {NoStop}%
\bibitem [{\citenamefont {Eder}(1980)}]{1980036}%
  \BibitemOpen
  \bibfield  {author} {\bibinfo {author} {\bibfnamefont {W.E.}\ \bibnamefont
  {Eder}},\ }\bibfield  {title} {\enquote {\bibinfo {title} {Rotary motion and
  SI},}\ }\href@noop {} {\bibfield  {journal} {\bibinfo  {journal} {Eur. J.
  Eng. Educ.}\ }\textbf {\bibinfo {volume} {4}},\ \bibinfo {pages} {319--327}
  (\bibinfo {year} {1980})}\BibitemShut {NoStop}%
\bibitem [{\citenamefont {Eder}(1982)}]{1982033}%
  \BibitemOpen
  \bibfield  {author} {\bibinfo {author} {\bibfnamefont {W.~E.}\ \bibnamefont
  {Eder}},\ }\bibfield  {title} {\enquote {\bibinfo {title} {A viewpoint on the
  quantity ``plane angle''},}\ }\href@noop {} {\bibfield  {journal} {\bibinfo
  {journal} {Metrologia}\ }\textbf {\bibinfo {volume} {18}},\ \bibinfo {pages}
  {1--12} (\bibinfo {year} {1982})}\BibitemShut {NoStop}%
\bibitem [{\citenamefont {Scott}(1985)}]{1985057}%
  \BibitemOpen
  \bibfield  {author} {\bibinfo {author} {\bibfnamefont {Bruce~L.}\
  \bibnamefont {Scott}},\ }\bibfield  {title} {\enquote {\bibinfo {title}
  {Letter to the editor},}\ }\href@noop {} {\bibfield  {journal} {\bibinfo
  {journal} {Am. J. Phys.}\ }\textbf {\bibinfo {volume} {53}},\ \bibinfo
  {pages} {520} (\bibinfo {year} {1985})}\BibitemShut {NoStop}%
\bibitem [{\citenamefont {Freeman}(1986)}]{1986059}%
  \BibitemOpen
  \bibfield  {author} {\bibinfo {author} {\bibfnamefont {G.~R.}\ \bibnamefont
  {Freeman}},\ }\bibfield  {title} {\enquote {\bibinfo {title} {Si units of
  frequency, angular velocity, planck's constant and $\hbar$},}\ }\href@noop {}
  {\bibfield  {journal} {\bibinfo  {journal} {Metrologia}\ }\textbf {\bibinfo
  {volume} {23}},\ \bibinfo {pages} {221--222} (\bibinfo {year}
  {1986})}\BibitemShut {NoStop}%
\bibitem [{\citenamefont {Torrens}(1986{\natexlab{a}})}]{1986056}%
  \BibitemOpen
  \bibfield  {author} {\bibinfo {author} {\bibfnamefont {A.~B.}\ \bibnamefont
  {Torrens}},\ }\bibfield  {title} {\enquote {\bibinfo {title} {On angles and
  angular quantities},}\ }\href@noop {} {\bibfield  {journal} {\bibinfo
  {journal} {Metrologia}\ }\textbf {\bibinfo {volume} {22}},\ \bibinfo {pages}
  {1--7} (\bibinfo {year} {1986}{\natexlab{a}})}\BibitemShut {NoStop}%
\bibitem [{\citenamefont {Thor}(1986)}]{1986058}%
  \BibitemOpen
  \bibfield  {author} {\bibinfo {author} {\bibfnamefont {A.~J.}\ \bibnamefont
  {Thor}},\ }\bibfield  {title} {\enquote {\bibinfo {title} {On angles and
  angular quantities},}\ }\href@noop {} {\bibfield  {journal} {\bibinfo
  {journal} {Metrologia}\ }\textbf {\bibinfo {volume} {23}},\ \bibinfo {pages}
  {55} (\bibinfo {year} {1986})}\BibitemShut {NoStop}%
\bibitem [{\citenamefont {Torrens}(1986{\natexlab{b}})}]{1986057}%
  \BibitemOpen
  \bibfield  {author} {\bibinfo {author} {\bibfnamefont {A.~B.}\ \bibnamefont
  {Torrens}},\ }\bibfield  {title} {\enquote {\bibinfo {title} {`on angles and
  angular quantities' Torrens' reply to Thor's comments},}\ }\href@noop {}
  {\bibfield  {journal} {\bibinfo  {journal} {Metrologia}\ }\textbf {\bibinfo
  {volume} {23}},\ \bibinfo {pages} {57--58} (\bibinfo {year}
  {1986}{\natexlab{b}})}\BibitemShut {NoStop}%
\bibitem [{\citenamefont {Wittmann}(1988)}]{1988061}%
  \BibitemOpen
  \bibfield  {author} {\bibinfo {author} {\bibfnamefont {H.}~\bibnamefont
  {Wittmann}},\ }\bibfield  {title} {\enquote {\bibinfo {title} {A new approach
  to the plane angle},}\ }\href@noop {} {\bibfield  {journal} {\bibinfo
  {journal} {Metrologia}\ }\textbf {\bibinfo {volume} {25}},\ \bibinfo {pages}
  {193--203} (\bibinfo {year} {1988})}\BibitemShut {NoStop}%
\bibitem [{\citenamefont {Brownstein}(1991)}]{1991100}%
  \BibitemOpen
  \bibfield  {author} {\bibinfo {author} {\bibfnamefont {K.~R.}\ \bibnamefont
  {Brownstein}},\ }\bibfield  {title} {\enquote {\bibinfo {title} {Concept of
  angle as a dimensional quantity},}\ }\href@noop {} {\bibfield  {journal}
  {\bibinfo  {journal} {AAPT}\ }\textbf {\bibinfo {volume} {21}},\ \bibinfo
  {pages} {92--93} (\bibinfo {year} {1991})}\BibitemShut {NoStop}%
\bibitem [{\citenamefont {Oberhofer}(1992)}]{1992087}%
  \BibitemOpen
  \bibfield  {author} {\bibinfo {author} {\bibfnamefont {E.~S.}\ \bibnamefont
  {Oberhofer}},\ }\bibfield  {title} {\enquote {\bibinfo {title} {What happens
  to the `radians'?}}\ }\href@noop {} {\bibfield  {journal} {\bibinfo
  {journal} {Phys. Teach.}\ }\textbf {\bibinfo {volume} {30}},\ \bibinfo
  {pages} {170--171} (\bibinfo {year} {1992})}\BibitemShut {NoStop}%
\bibitem [{\citenamefont {French}(1992)}]{1992088}%
  \BibitemOpen
  \bibfield  {author} {\bibinfo {author} {\bibfnamefont {Anthony~P.}\
  \bibnamefont {French}},\ }\bibfield  {title} {\enquote {\bibinfo {title}
  {What happens to the `radians'?}}\ }\href@noop {} {\bibfield  {journal}
  {\bibinfo  {journal} {Phys. Teach.}\ }\textbf {\bibinfo {volume} {30}},\
  \bibinfo {pages} {260--261} (\bibinfo {year} {1992})}\BibitemShut {NoStop}%
\bibitem [{\citenamefont {Cooper}(1992)}]{1992089}%
  \BibitemOpen
  \bibfield  {author} {\bibinfo {author} {\bibfnamefont {Michael}\ \bibnamefont
  {Cooper}},\ }\bibfield  {title} {\enquote {\bibinfo {title} {Who named the
  radian?}}\ }\href@noop {} {\bibfield  {journal} {\bibinfo  {journal} {Math.
  Gaz.}\ }\textbf {\bibinfo {volume} {76}},\ \bibinfo {pages} {100--101}
  (\bibinfo {year} {1992})}\BibitemShut {NoStop}%
\bibitem [{\citenamefont {Aubrecht}\ \emph {et~al.}(1993)\citenamefont
  {Aubrecht}, \citenamefont {French}, \citenamefont {Iona},\ and\ \citenamefont
  {Welch}}]{1993136}%
  \BibitemOpen
  \bibfield  {author} {\bibinfo {author} {\bibfnamefont {Gordon~J.}\
  \bibnamefont {Aubrecht}, \bibfnamefont {II}}, \bibinfo {author}
  {\bibfnamefont {Anthony~P.}\ \bibnamefont {French}}, \bibinfo {author}
  {\bibfnamefont {Mario}\ \bibnamefont {Iona}}, \ and\ \bibinfo {author}
  {\bibfnamefont {Daniel~W.}\ \bibnamefont {Welch}},\ }\bibfield  {title}
  {\enquote {\bibinfo {title} {The radian - that troublesome unit},}\
  }\href@noop {} {\bibfield  {journal} {\bibinfo  {journal} {Phys. Teach.}\
  }\textbf {\bibinfo {volume} {31}},\ \bibinfo {pages} {84--87} (\bibinfo
  {year} {1993})}\BibitemShut {NoStop}%
\bibitem [{\citenamefont {Scott}(1993)}]{1993137}%
  \BibitemOpen
  \bibfield  {author} {\bibinfo {author} {\bibfnamefont {B.~L.}\ \bibnamefont
  {Scott}},\ }\bibfield  {title} {\enquote {\bibinfo {title} {It's obvious
  -now},}\ }\href@noop {} {\bibfield  {journal} {\bibinfo  {journal} {Phys.
  Teach.}\ }\textbf {\bibinfo {volume} {31}},\ \bibinfo {pages} {262} (\bibinfo
  {year} {1993})}\BibitemShut {NoStop}%
\bibitem [{\citenamefont {Brownstein}(1997)}]{1997199}%
  \BibitemOpen
  \bibfield  {author} {\bibinfo {author} {\bibfnamefont {K.~R.}\ \bibnamefont
  {Brownstein}},\ }\bibfield  {title} {\enquote {\bibinfo {title}
  {Angles$-$let's treat them squarely},}\ }\href@noop {} {\bibfield  {journal}
  {\bibinfo  {journal} {Am. J. Phys.}\ }\textbf {\bibinfo {volume} {65}},\
  \bibinfo {pages} {605--614} (\bibinfo {year} {1997})}\BibitemShut {NoStop}%
\bibitem [{\citenamefont {L\'evy-Leblond}(1998)}]{1998167}%
  \BibitemOpen
  \bibfield  {author} {\bibinfo {author} {\bibfnamefont {Jean-Marc}\
  \bibnamefont {L\'evy-Leblond}},\ }\bibfield  {title} {\enquote {\bibinfo
  {title} {Dimensional angles and universal constants},}\ }\href@noop {}
  {\bibfield  {journal} {\bibinfo  {journal} {Am. J. Phys.}\ }\textbf {\bibinfo
  {volume} {66}},\ \bibinfo {pages} {814--815} (\bibinfo {year}
  {1998})}\BibitemShut {NoStop}%
\bibitem [{\citenamefont {Mills}\ \emph {et~al.}(2001)\citenamefont {Mills},
  \citenamefont {Taylor},\ and\ \citenamefont {Thor}}]{2001385}%
  \BibitemOpen
  \bibfield  {author} {\bibinfo {author} {\bibfnamefont {I.~M.}\ \bibnamefont
  {Mills}}, \bibinfo {author} {\bibfnamefont {B.~N.}\ \bibnamefont {Taylor}}, \
  and\ \bibinfo {author} {\bibfnamefont {A.~J.}\ \bibnamefont {Thor}},\
  }\bibfield  {title} {\enquote {\bibinfo {title} {Definitions of the units
  radian, neper, bel and decibel},}\ }\href@noop {} {\bibfield  {journal}
  {\bibinfo  {journal} {Metrologia}\ }\textbf {\bibinfo {volume} {38}},\
  \bibinfo {pages} {353--361} (\bibinfo {year} {2001})}\BibitemShut {NoStop}%
\bibitem [{\citenamefont {Emerson}(2002)}]{2002280}%
  \BibitemOpen
  \bibfield  {author} {\bibinfo {author} {\bibfnamefont {W.~H.}\ \bibnamefont
  {Emerson}},\ }\bibfield  {title} {\enquote {\bibinfo {title} {A reply to
  ``definitions of the units radian, neper, bel and decibel'' by I. M. Mills et
  al.}}\ }\href@noop {} {\bibfield  {journal} {\bibinfo  {journal}
  {Metrologia}\ }\textbf {\bibinfo {volume} {39}},\ \bibinfo {pages} {105--109}
  (\bibinfo {year} {2002})}\BibitemShut {NoStop}%
\bibitem [{\citenamefont {Emerson}(2005{\natexlab{a}})}]{2005364}%
  \BibitemOpen
  \bibfield  {author} {\bibinfo {author} {\bibfnamefont {W.~H.}\ \bibnamefont
  {Emerson}},\ }\bibfield  {title} {\enquote {\bibinfo {title} {On the concept
  of $dimension$},}\ }\href@noop {} {\bibfield  {journal} {\bibinfo  {journal}
  {Metrologia}\ }\textbf {\bibinfo {volume} {42}},\ \bibinfo {pages} {L21--L22}
  (\bibinfo {year} {2005}{\natexlab{a}})}\BibitemShut {NoStop}%
\bibitem [{\citenamefont {Emerson}(2005{\natexlab{b}})}]{2005363}%
  \BibitemOpen
  \bibfield  {author} {\bibinfo {author} {\bibfnamefont {W.~H.}\ \bibnamefont
  {Emerson}},\ }\bibfield  {title} {\enquote {\bibinfo {title} {Differing
  angles on $angle$},}\ }\href@noop {} {\bibfield  {journal} {\bibinfo
  {journal} {Metrologia}\ }\textbf {\bibinfo {volume} {42}},\ \bibinfo {pages}
  {L23--L26} (\bibinfo {year} {2005}{\natexlab{b}})}\BibitemShut {NoStop}%
\bibitem [{\citenamefont {Karshenboim}(2005)}]{2005109}%
  \BibitemOpen
  \bibfield  {author} {\bibinfo {author} {\bibfnamefont {Savely~G.}\
  \bibnamefont {Karshenboim}},\ }\bibfield  {title} {\enquote {\bibinfo {title}
  {Fundmental physical constants: looking from different angles},}\ }\href@noop
  {} {\bibfield  {journal} {\bibinfo  {journal} {Can. J. Phys.}\ }\textbf
  {\bibinfo {volume} {83}},\ \bibinfo {pages} {767--811} (\bibinfo {year}
  {2005})}\BibitemShut {NoStop}%
\bibitem [{\citenamefont {Foster}(2010)}]{2010207}%
  \BibitemOpen
  \bibfield  {author} {\bibinfo {author} {\bibfnamefont {Marcus~P.}\
  \bibnamefont {Foster}},\ }\bibfield  {title} {\enquote {\bibinfo {title} {The
  next 50 years of the SI: a review of the opportunities for the e-science
  age},}\ }\href@noop {} {\bibfield  {journal} {\bibinfo  {journal}
  {Metrologia}\ }\textbf {\bibinfo {volume} {47}},\ \bibinfo {pages} {R41--R51}
  (\bibinfo {year} {2010})}\BibitemShut {NoStop}%
\bibitem [{\citenamefont {Mohr}\ and\ \citenamefont
  {Phillips}(2015{\natexlab{a}})}]{2015004}%
  \BibitemOpen
  \bibfield  {author} {\bibinfo {author} {\bibfnamefont {Peter~J.}\
  \bibnamefont {Mohr}}\ and\ \bibinfo {author} {\bibfnamefont {William~D.}\
  \bibnamefont {Phillips}},\ }\bibfield  {title} {\enquote {\bibinfo {title}
  {Dimensionless units in the SI},}\ }\href@noop {} {\bibfield  {journal}
  {\bibinfo  {journal} {Metrologia}\ }\textbf {\bibinfo {volume} {52}},\
  \bibinfo {pages} {40--47} (\bibinfo {year} {2015}{\natexlab{a}})}\BibitemShut
  {NoStop}%
\bibitem [{\citenamefont {Leonard}(2015)}]{2015048}%
  \BibitemOpen
  \bibfield  {author} {\bibinfo {author} {\bibfnamefont {B.~P.}\ \bibnamefont
  {Leonard}},\ }\bibfield  {title} {\enquote {\bibinfo {title} {Comment on
  `Dimensionless units in the SI'},}\ }\href@noop {} {\bibfield  {journal}
  {\bibinfo  {journal} {Metrologia}\ }\textbf {\bibinfo {volume} {52}},\
  \bibinfo {pages} {613--616} (\bibinfo {year} {2015})}\BibitemShut {NoStop}%
\bibitem [{\citenamefont {Mohr}\ and\ \citenamefont
  {Phillips}(2015{\natexlab{b}})}]{2015049}%
  \BibitemOpen
  \bibfield  {author} {\bibinfo {author} {\bibfnamefont {Peter~J.}\
  \bibnamefont {Mohr}}\ and\ \bibinfo {author} {\bibfnamefont {Williams~D.}\
  \bibnamefont {Phillips}},\ }\bibfield  {title} {\enquote {\bibinfo {title}
  {Reply to comments on `Dimensionless units in the SI'},}\ }\href@noop {}
  {\bibfield  {journal} {\bibinfo  {journal} {Metrologia}\ }\textbf {\bibinfo
  {volume} {52}},\ \bibinfo {pages} {617--618} (\bibinfo {year}
  {2015}{\natexlab{b}})}\BibitemShut {NoStop}%
\bibitem [{\citenamefont {Krystek}(2015)}]{2015154}%
  \BibitemOpen
  \bibfield  {author} {\bibinfo {author} {\bibfnamefont {M.~P.}\ \bibnamefont
  {Krystek}},\ }\bibfield  {title} {\enquote {\bibinfo {title} {The term
  `dimension' in the International System of Units},}\ }\href@noop {}
  {\bibfield  {journal} {\bibinfo  {journal} {Metrologia}\ }\textbf {\bibinfo
  {volume} {52}},\ \bibinfo {pages} {297--300} (\bibinfo {year}
  {2015})}\BibitemShut {NoStop}%
\bibitem [{\citenamefont {Quincey}(2016{\natexlab{a}})}]{2016021}%
  \BibitemOpen
  \bibfield  {author} {\bibinfo {author} {\bibfnamefont {Paul}\ \bibnamefont
  {Quincey}},\ }\bibfield  {title} {\enquote {\bibinfo {title} {The range of
  options for handling plane angle and solid angle within a system of units},}\
  }\href@noop {} {\bibfield  {journal} {\bibinfo  {journal} {Metrologia}\
  }\textbf {\bibinfo {volume} {53}},\ \bibinfo {pages} {840--845} (\bibinfo
  {year} {2016}{\natexlab{a}})}\BibitemShut {NoStop}%
\bibitem [{\citenamefont {Mills}(2016{\natexlab{a}})}]{2016022}%
  \BibitemOpen
  \bibfield  {author} {\bibinfo {author} {\bibfnamefont {Ian}\ \bibnamefont
  {Mills}},\ }\bibfield  {title} {\enquote {\bibinfo {title} {On the units
  radian and cycle for the quantity plane angle},}\ }\href@noop {} {\bibfield
  {journal} {\bibinfo  {journal} {Metrologia}\ }\textbf {\bibinfo {volume}
  {53}},\ \bibinfo {pages} {991--997} (\bibinfo {year}
  {2016}{\natexlab{a}})}\BibitemShut {NoStop}%
\bibitem [{\citenamefont {Leonard}(2016)}]{2016059}%
  \BibitemOpen
  \bibfield  {author} {\bibinfo {author} {\bibfnamefont {B.~P.}\ \bibnamefont
  {Leonard}},\ }\bibfield  {title} {\enquote {\bibinfo {title} {Comment on `On
  the units radian and cycle for the quantity plane angle'},}\ }\href@noop {}
  {\bibfield  {journal} {\bibinfo  {journal} {Metrologia}\ }\textbf {\bibinfo
  {volume} {53}},\ \bibinfo {pages} {1281--1285} (\bibinfo {year}
  {2016})}\BibitemShut {NoStop}%
\bibitem [{\citenamefont {Mills}(2016{\natexlab{b}})}]{2016060}%
  \BibitemOpen
  \bibfield  {author} {\bibinfo {author} {\bibfnamefont {I.~M.}\ \bibnamefont
  {Mills}},\ }\bibfield  {title} {\enquote {\bibinfo {title} {Reply to comment
  on 'On the units radian and cycle for the quantity plane angle'},}\
  }\href@noop {} {\bibfield  {journal} {\bibinfo  {journal} {Metrologia}\
  }\textbf {\bibinfo {volume} {53}},\ \bibinfo {pages} {1286--0} (\bibinfo
  {year} {2016}{\natexlab{b}})}\BibitemShut {NoStop}%
\bibitem [{\citenamefont {Quincey}\ and\ \citenamefont
  {Brown}(2016)}]{2016023}%
  \BibitemOpen
  \bibfield  {author} {\bibinfo {author} {\bibfnamefont {Paul}\ \bibnamefont
  {Quincey}}\ and\ \bibinfo {author} {\bibfnamefont {Richard J.~C.}\
  \bibnamefont {Brown}},\ }\bibfield  {title} {\enquote {\bibinfo {title}
  {Implications of adopting plane angle as a base quantity in the SI},}\
  }\href@noop {} {\bibfield  {journal} {\bibinfo  {journal} {Metrologia}\
  }\textbf {\bibinfo {volume} {53}},\ \bibinfo {pages} {998--1002} (\bibinfo
  {year} {2016})}\BibitemShut {NoStop}%
\bibitem [{\citenamefont {Quincey}(2016{\natexlab{b}})}]{2016121}%
  \BibitemOpen
  \bibfield  {author} {\bibinfo {author} {\bibfnamefont {Paul}\ \bibnamefont
  {Quincey}},\ }\bibfield  {title} {\enquote {\bibinfo {title} {Natural units
  in physics, and the curious case of the radian},}\ }\href@noop {} {\bibfield
  {journal} {\bibinfo  {journal} {Phys. Educ.}\ }\textbf {\bibinfo {volume}
  {51}},\ \bibinfo {eid} {065012} (\bibinfo {year}
  {2016}{\natexlab{b}})}\BibitemShut {NoStop}%
\bibitem [{\citenamefont {Editorials}(2017)}]{2017130}%
  \BibitemOpen
  \bibfield  {author} {\bibinfo {author} {\bibfnamefont {Nature}\ \bibnamefont
  {Editorials}},\ }\bibfield  {title} {\enquote {\bibinfo {title} {Lost
  dimension},}\ }\href@noop {} {\bibfield  {journal} {\bibinfo  {journal}
  {Nature}\ }\textbf {\bibinfo {volume} {548}},\ \bibinfo {pages} {135}
  (\bibinfo {year} {2017})}\BibitemShut {NoStop}%
\bibitem [{\citenamefont {Flater}(2017)}]{2017154}%
  \BibitemOpen
  \bibfield  {author} {\bibinfo {author} {\bibfnamefont {David}\ \bibnamefont
  {Flater}},\ }\bibfield  {title} {\enquote {\bibinfo {title} {Redressing
  grievances with the treatment of dimensionless quantities in SI},}\
  }\href@noop {} {\bibfield  {journal} {\bibinfo  {journal} {Measurement}\
  }\textbf {\bibinfo {volume} {109}},\ \bibinfo {pages} {105--110} (\bibinfo
  {year} {2017})}\BibitemShut {NoStop}%
\bibitem [{\citenamefont {Bich}(2019)}]{2019004}%
  \BibitemOpen
  \bibfield  {author} {\bibinfo {author} {\bibfnamefont {W.}~\bibnamefont
  {Bich}},\ }\bibfield  {title} {\enquote {\bibinfo {title} {The
  third-millennium International System of Units},}\ }\href@noop {} {\bibfield
  {journal} {\bibinfo  {journal} {Riv. Nuovo Cimento}\ }\textbf {\bibinfo
  {volume} {42}},\ \bibinfo {pages} {49--102} (\bibinfo {year}
  {2019})}\BibitemShut {NoStop}%
\bibitem [{\citenamefont {Kalinin}(2019)}]{2019108}%
  \BibitemOpen
  \bibfield  {author} {\bibinfo {author} {\bibfnamefont {M.~I.}\ \bibnamefont
  {Kalinin}},\ }\bibfield  {title} {\enquote {\bibinfo {title} {On the status
  of plane and solid angles in the International System of Units (SI)},}\
  }\href@noop {} {\bibfield  {journal} {\bibinfo  {journal} {Metrologia}\
  }\textbf {\bibinfo {volume} {56}},\ \bibinfo {eid} {065009} (\bibinfo {year}
  {2019})}\BibitemShut {NoStop}%
\bibitem [{\citenamefont {Quincey}\ and\ \citenamefont
  {Burrows}(2019)}]{2019112}%
  \BibitemOpen
  \bibfield  {author} {\bibinfo {author} {\bibfnamefont {Paul}\ \bibnamefont
  {Quincey}}\ and\ \bibinfo {author} {\bibfnamefont {Kathryn}\ \bibnamefont
  {Burrows}},\ }\bibfield  {title} {\enquote {\bibinfo {title} {The role of
  unit systems in expressing and testing the laws of nature},}\ }\href@noop {}
  {\bibfield  {journal} {\bibinfo  {journal} {Metrologia}\ }\textbf {\bibinfo
  {volume} {56}},\ \bibinfo {eid} {065001} (\bibinfo {year}
  {2019})}\BibitemShut {NoStop}%
\bibitem [{\citenamefont {Bunker}\ \emph {et~al.}(2019)\citenamefont {Bunker},
  \citenamefont {Mills},\ and\ \citenamefont {Jensen}}]{2019117}%
  \BibitemOpen
  \bibfield  {author} {\bibinfo {author} {\bibfnamefont {P.~R.}\ \bibnamefont
  {Bunker}}, \bibinfo {author} {\bibfnamefont {Ian~M.}\ \bibnamefont {Mills}},
  \ and\ \bibinfo {author} {\bibfnamefont {Per}\ \bibnamefont {Jensen}},\
  }\bibfield  {title} {\enquote {\bibinfo {title} {The Planck constant and its
  units},}\ }\href@noop {} {\bibfield  {journal} {\bibinfo  {journal} {J.
  Quant. Spectrosc. Radiat. Transfer}\ }\textbf {\bibinfo {volume} {237}},\
  \bibinfo {eid} {106594} (\bibinfo {year} {2019})}\BibitemShut {NoStop}%
\bibitem [{\citenamefont {Lovatt}(2019)}]{2019121}%
  \BibitemOpen
  \bibfield  {author} {\bibinfo {author} {\bibfnamefont {Ian}\ \bibnamefont
  {Lovatt}},\ }\bibfield  {title} {\enquote {\bibinfo {title} {The physics of
  ${\lim}_{\theta\rightarrow0}$(sin $\theta$)/$\theta$ = 1},}\ }\href@noop {}
  {\bibfield  {journal} {\bibinfo  {journal} {Phys. Teach.}\ }\textbf {\bibinfo
  {volume} {57}},\ \bibinfo {pages} {558--559} (\bibinfo {year}
  {2019})}\BibitemShut {NoStop}%
\bibitem [{\citenamefont {Bunker}\ and\ \citenamefont
  {Jensen}(2020)}]{2020031}%
  \BibitemOpen
  \bibfield  {author} {\bibinfo {author} {\bibfnamefont {P.~R.}\ \bibnamefont
  {Bunker}}\ and\ \bibinfo {author} {\bibfnamefont {Per}\ \bibnamefont
  {Jensen}},\ }\bibfield  {title} {\enquote {\bibinfo {title} {The planck
  constant of action $h_{\rm a}$},}\ }\href@noop {} {\bibfield  {journal} {\bibinfo
  {journal} {J. Quant. Spectrosc. Radiat. Transfer}\ }\textbf {\bibinfo
  {volume} {243}},\ \bibinfo {eid} {106835} (\bibinfo {year}
  {2020})}\BibitemShut {NoStop}%
\bibitem [{\citenamefont {Quincey}(2020)}]{2020078}%
  \BibitemOpen
  \bibfield  {author} {\bibinfo {author} {\bibfnamefont {Paul}\ \bibnamefont
  {Quincey}},\ }\bibfield  {title} {\enquote {\bibinfo {title} {Angles in the
  SI: treating the radian as an independent, unhidden unit does not require the
  redefinition of the term `frequency' or the unit hertz},}\ }\href@noop {}
  {\bibfield  {journal} {\bibinfo  {journal} {Metrologia}\ }\textbf {\bibinfo
  {volume} {57}},\ \bibinfo {eid} {053001} (\bibinfo {year}
  {2020})}\BibitemShut {NoStop}%
\bibitem [{\citenamefont {Leonard}(2021)}]{2021025}%
  \BibitemOpen
  \bibfield  {author} {\bibinfo {author} {\bibfnamefont {B.~P.}\ \bibnamefont
  {Leonard}},\ }\bibfield  {title} {\enquote {\bibinfo {title} {Proposal for
  the dimensionally consistent treatment of angle and solid angle by the
  International System of Units (SI)},}\ }\href@noop {} {\bibfield  {journal}
  {\bibinfo  {journal} {Metrologia}\ }\textbf {\bibinfo {volume} {58}},\
  \bibinfo {eid} {052001} (\bibinfo {year} {2021})}\BibitemShut {NoStop}%
\bibitem [{\citenamefont {Quincey}(2021)}]{2021029}%
  \BibitemOpen
  \bibfield  {author} {\bibinfo {author} {\bibfnamefont {Paul}\ \bibnamefont
  {Quincey}},\ }\bibfield  {title} {\enquote {\bibinfo {title} {Angles in the
  SI: a detailed proposal for solving the problem},}\ }\href@noop {} {\bibfield
   {journal} {\bibinfo  {journal} {Metrologia}\ }\textbf {\bibinfo {volume}
  {58}},\ \bibinfo {eid} {053002} (\bibinfo {year} {2021})}\BibitemShut
  {NoStop}%
\bibitem [{mat()}]{math}%
  \BibitemOpen
  \href@noop {} {}\bibinfo {howpublished} {Mention of specific commercial
  products does not imply endorsement by NIST.}\BibitemShut {Stop}%
\bibitem [{\citenamefont {Buckingham}(1914)}]{1914002}%
  \BibitemOpen
  \bibfield  {author} {\bibinfo {author} {\bibfnamefont {E.}~\bibnamefont
  {Buckingham}},\ }\bibfield  {title} {\enquote {\bibinfo {title} {On
  physically similar systems; illustrations of the use of dimensional
  equations},}\ }\href@noop {} {\bibfield  {journal} {\bibinfo  {journal}
  {Phys. Rev.}\ }\textbf {\bibinfo {volume} {4}},\ \bibinfo {pages} {345--376}
  (\bibinfo {year} {1914})}\BibitemShut {NoStop}%
\bibitem [{err()}]{error}%
  \BibitemOpen
  \href@noop {} {}\bibinfo {howpublished} {The ``rule of thumb'' unit for
  centripital force given in Table II of \citet{2015004} is not consistent with
  this result.}\BibitemShut {Stop}%
\bibitem [{wol(2022)}]{wolf}%
  \BibitemOpen
  \href@noop {} {}\bibinfo {howpublished}
  {https://mathworld.wolfram.com/JacobiEllipticFunctions.html} (\bibinfo {year}
  {2022})\BibitemShut {NoStop}%
\bibitem [{\citenamefont {BIPM}(2019)}]{2019SI}%
  \BibitemOpen
  \bibfield  {author} {\bibinfo {author} {\bibnamefont {BIPM}},\ }\href@noop {}
  {\emph {\bibinfo {title} {Le Syst\`eme international d'unit\'es (SI)}}},\
  \bibinfo {edition} {9th}\ ed.\ (\bibinfo  {publisher} {Bureau International
  des Poids et Mesures},\ \bibinfo {address} {S\`evres, France},\ \bibinfo
  {year} {2019})\BibitemShut {NoStop}%
\bibitem [{\citenamefont {Jackson}(1998)}]{1998165}%
  \BibitemOpen
  \bibfield  {author} {\bibinfo {author} {\bibfnamefont {John~David}\
  \bibnamefont {Jackson}},\ }\href@noop {} {\emph {\bibinfo {title} {Classical
  Electrodynamics}}},\ \bibinfo {edition} {3rd}\ ed.\ (\bibinfo  {publisher}
  {John Wiley and Sons, Inc},\ \bibinfo {address} {New York, NY},\ \bibinfo
  {year} {1998}), see p. 775 \BibitemShut {NoStop}%
\end{thebibliography}%
\end{document}